\documentclass[10pt,letterpaper]{article}
\usepackage{opex3}
\usepackage{cite}
\usepackage{wasysym}

\begin{document}




\title{Wavelength dependence of electron localization in the laser-driven dissociation of H$_2^+$}

\author{Kunlong Liu,$^{1}$ Weiyi Hong,$^{1}$ Qingbin Zhang,$^{1}$ and Peixiang Lu$^{1,2}$$^*$}

\address{$^1$Wuhan National Laboratory for Optoelectronics, Huazhong University of Science and Technology, Wuhan 430074, China \\
$^2$Laboratory of Optical Information Technology, Wuhan Institute of Technology, Wuhan 430073, China }
\email{$^*$lupeixiang@mail.hust.edu.cn} 

\begin{abstract}
We theoretically investigate the laser wavelength dependence of asymmetric dissociation of H$_2^+$. It is found that the electron localization in molecular dissociation is significantly manipulated by varying the wavelength of the driving field. Through creating a strong nuclear vibration in the laser-molecular interaction, our simulations demonstrate that the few-cycle mid-infrared pulse can effectively localize the electron at one of the dissociating nuclei with weak ionization. Moreover, we show that the observed phase-shift of the dissociation asymmetry is attributed to the different population transfers by the remaining fields after the internuclear distances reach the one-photon coupling point.
\end{abstract}


\section{Introduction}

The interaction of intense laser pulses with atoms and molecules has been a very important topic in current research of attosecond physics. Recent advances in femtosecond (fs) laser technology have made it possible for scientists to observe, study and control the ultrafast molecular dynamics in the strong field \cite{Krausz}. Several mechanisms responsible for the laser induced dynamics of the molecule have been discovered in theoretical and experimental studies on H$_2$ and its isotopes, such as bond softening, bond hardening, above threshold dissociation and so on \cite{Posthumus}.

The electron localization in the dissociation of molecules has recently become an interesting subject for molecular research \cite{Bandrauk2004, Roudnev}.
It is crucial for steering many physical and chemical processes to control the electron localization in molecules on the attosecond time-scale \cite{Krausz}.
A landmark experimental demonstration in this direction was achieved by Kling \textit{et al.} \cite{Kling}, where a few-cycle carrier-envelope phase (CEP) stabilized pulse was used to control the localization of the electron in the dissociation of D$_2^+$.
As stimulated by the pioneer works \cite{Bandrauk2004, Roudnev,Kling}, many methods have been proposed to control as well as explain the electron localization in small molecules \cite{Tong,Grafe,He2,
Kremer,Ray,Sansone,Singh,Fischer,Calvert,Freek} and  heteronuclear molecules \cite{Znakovskaya,Liu}.
However, partly due to the limitation imposed by ultrafast amplifier technology, the wavelengths of the controlling pulse in those previous studies are mainly in the near-infrared region.
Nonetheless, it has been shown that the dynamics of the nuclear wavepackets, which can be used to steer the localization of the electron in molecules \cite{Fischer}, are significantly changed by mid-infrared pulses \cite{Liucd,MM}.
It is therefore appealing to know how the mid-infrared pulses influence the asymmetry of electron localization in molecular dissociation.
Moreover, with the development of the infrared optical parametric amplifier system \cite{Gu}, the mid-infrared pulses have been widely utilized to study as well as control the electron dynamics in the ultrafast interaction \cite{Tate,Zhang,Vozzi,Hong}. The understanding of wavelength effect on the electron localization in molecular dissociation will provide helpful experimental guidance on steering physical and chemical processes in the future.

In the present work, the dependence of the asymmetric dissociation of H$_2^+$ on the wavelength of the driving field is studied by numerical solutions of a non-Born-Oppenheimer time-dependent Schr$\mathrm{\ddot{o}}$dinger equation.
The results show that both the amplitude and the relative phase of the CEP-sensitive asymmetry are changed significantly with the wavelength. We then qualitatively explain the wavelength-dependent dissociation asymmetry by analyzing the real-time propagation of the electron-nuclear wavepackets during the interaction.
Further results demonstrate
that
the few-cycle mid-infrared pulse can create a strong nuclear vibration and effectively localize the electron at one of the dissociating nuclei with weak ionization. Moreover, the observed phase-shift of the dissociation asymmetry is found to be attributed to the different population transfers by the remaining fields after the internuclear distances reach the one-photon coupling point.
Although the recollision induced dissociation is demonstrated to be an important dissociation channel \cite{Kling,Tong,Grafe} and also expected to be wavelength dependent, it is not considered in the present study.

\section{Theoretical model}
For numerical simulation we have solved the non-Born-Oppenheimer time-dependent Schr$\ddot{\mathrm{o}}$dinger equation (TDSE) for a reduced-dimensionality model of H$_2^+$. The model consists of one-dimensional motion of the nuclei and one-dimensional motion of the electron \cite{Kulander,Steeg}. In this model, the electronic and nuclear motions are restricted along the polarization direction of the linearly polarized laser pulse.
The time-dependent Schr$\ddot{\mathrm{o}}$dinger equation is written as (Hartree atomic units are used throughout unless otherwise indicated):
\begin{eqnarray}
i\frac{\partial}{\partial t}\Psi(R,z;t)=[H_0+V(t)]\Psi(R,z;t),
\end{eqnarray}
where $H_0$ is the field-free Hamiltonian,
\begin{eqnarray}
H_0=-\frac{1}{2\mu}\frac{\partial ^2}{\partial R^2}
-\frac{1}{2\mu _e}\frac{\partial ^2}{\partial z^2}
-\frac{1}{\sqrt{(z-R/2)^2+\alpha}}
-\frac{1}{\sqrt{(z+R/2)^2+\alpha}}
+\frac{1}{\sqrt{R^2+\beta}},
\end{eqnarray}
and $V(t)$ is the electric potential including the laser-molecular interaction. Here, $R$ is the internuclear distance, $z$ is the electron position measured from the center-of-mass of the protons, $\mu=m_p/2$ and $\mu_e=2m_p/(2m_p+1)$ are the reduced masses with $m_p$ the mass of the proton, and $\alpha=1$ and $\beta=0.03$ are the soft-core parameters. The interaction with the laser is given by (in the dipole approximation and length gauge) \cite{Hiskes} $V(t)=[1+(2m_p+1)^{-1}]z E(t)$
with $E(t)=E_0 \exp[-2\ln(2)(t/\tau)^{2}]\sin (\omega t+\phi)$. $E_{0}$ is the peak electric field amplitude, $\tau$ is the pulse duration, $\omega$ is the central frequency, and $\phi$ is the CEP of the pulse. The electric field is chosen to be linearly polarized along the $z$ axis. Then one can solve Eq.(1) by using the Crank-Nicolson method. Compared to the 3D TDSE methods used in Refs. \cite{Roudnev,He2,Liu}, the electric motion is restricted along the $z$ direction for the present approach, and consequently the electronic wavepackets spread in the radial direction (perpendicular to the $z$-axis) is not taken into account. But it has been shown before that this type of reduced-dimensional model reproduces experimental result at least qualitatively \cite{Kulander,Steeg}.

\begin{figure}
\centering\includegraphics[width=12.8cm]{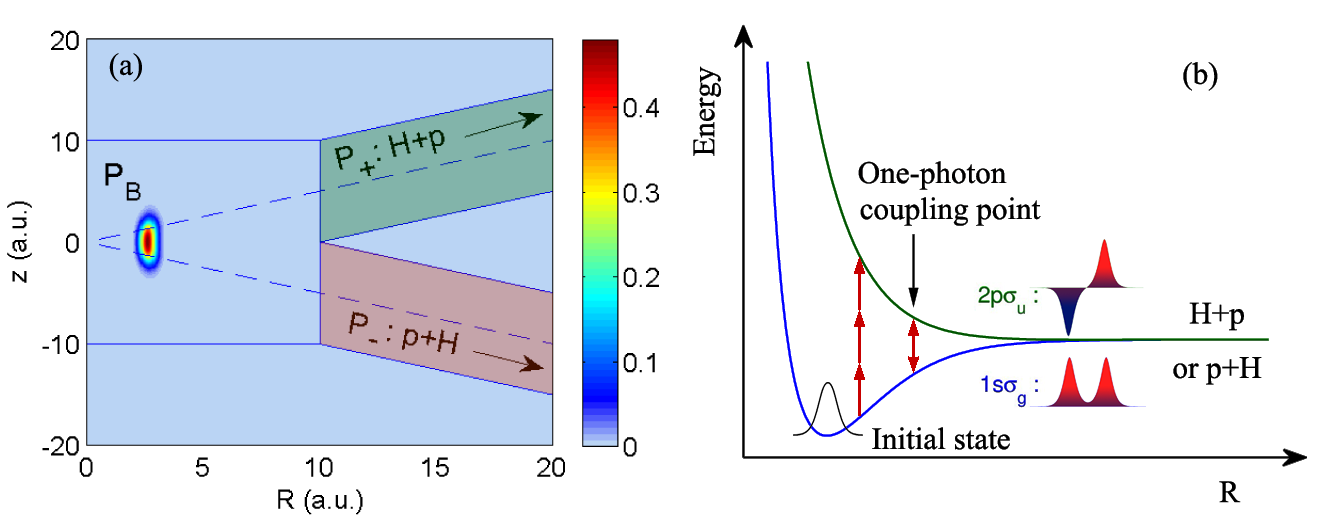}
\caption{\label{fig1} (a) The probability density of the H$_2^+$ molecular ground state and the configuration space regions associated with different final state channels. The dash lines indicate the positions of the nuclei. The color scale is linear. (b) The sketch of the potential energy surface for the $1s\sigma_g$ and $2p\sigma_u$ states of H$_2^+$.}
\end{figure}

In the present simulation, the wave function is discrete on a two-dimensional grid which ranges in $R$-direction from 0 to 32 a.u., for $z$ from -300 to 300 a.u. with 800 and 3000 points, respectively.
We have defined different regions of the grid to distinguish the dissociation, ionization and bound probabilities.
As illustrated in Fig.~\ref{fig1}(a), the bound region is defined as
\begin{eqnarray}
P_\mathrm{B}:\ 0<R<10 \quad \mathrm{and} \quad -10<z<10,
\end{eqnarray}
and the two channels of dissociation are defined as
\begin{eqnarray}
P_+:\ R>10 \quad \mathrm{and} \quad \sqrt{(z-R/2)^2}<5, \\
P_-:\ R>10 \quad \mathrm{and} \quad \sqrt{(z+R/2)^2}<5,
\end{eqnarray}
where $P_+$ and $P_-$ indicate the probabilities of the electron localization on the upper and lower protons ($p$), respectively, and thus these are the directions that the H atom is emitted.
By integrating the density of the final wavepackets in the regions described above, the final localization and bound probabilities can be obtained.
The region describing the ionization is not contained in Fig.~\ref{fig1} but we estimate the ionization probability by storing the absorbed wavepackets reaching the boundary in the $z$ direction.
In our calculation, the ground state of H$_2^+$ (electronic $1s\sigma_g$ state and vibrational $v=0$ state), as shown in Fig.~\ref{fig1}(a), is chosen to be the initial state of the system and obtained by propagating the field-free Schr$\ddot{\mathrm{o}}$dinger equation in imaginary time.
The evolution of the wave function in the external field has been continued with a time step of $\delta t=0.05$ until the probability of the bound region is converged. During the propagation, we have employed absorbing boundaries using cos$^{1/6}$-masking functions but stored the absorbed contributions as dissociation or ionization, respectively.

For better understanding of our simulation, here we briefly introduce the origin of the electron localization asymmetry in molecular dissociation. Figure~\ref{fig1}(b) shows the sketch of the potential energy surface for the $1s\sigma_g$ and $2p\sigma_u$ states of H$_2^+$. These two states are the gerade and ungerade states, respectively, and the superposition of the dissociative wavepackets on the two states will lead to the electron localization on one of the nuclei. Depending on the CEP \cite{Roudnev}, intensity \cite{He2} and even wavelength of the pulse, population is transferred between the two states, changing the electron localization between the nuclei. The detail of the physical mechanism responsible for the asymmetry of electron localization will be discussed in Section 3.

\section{Results and discussion}

\begin{figure}
\centering\includegraphics[width=11cm]{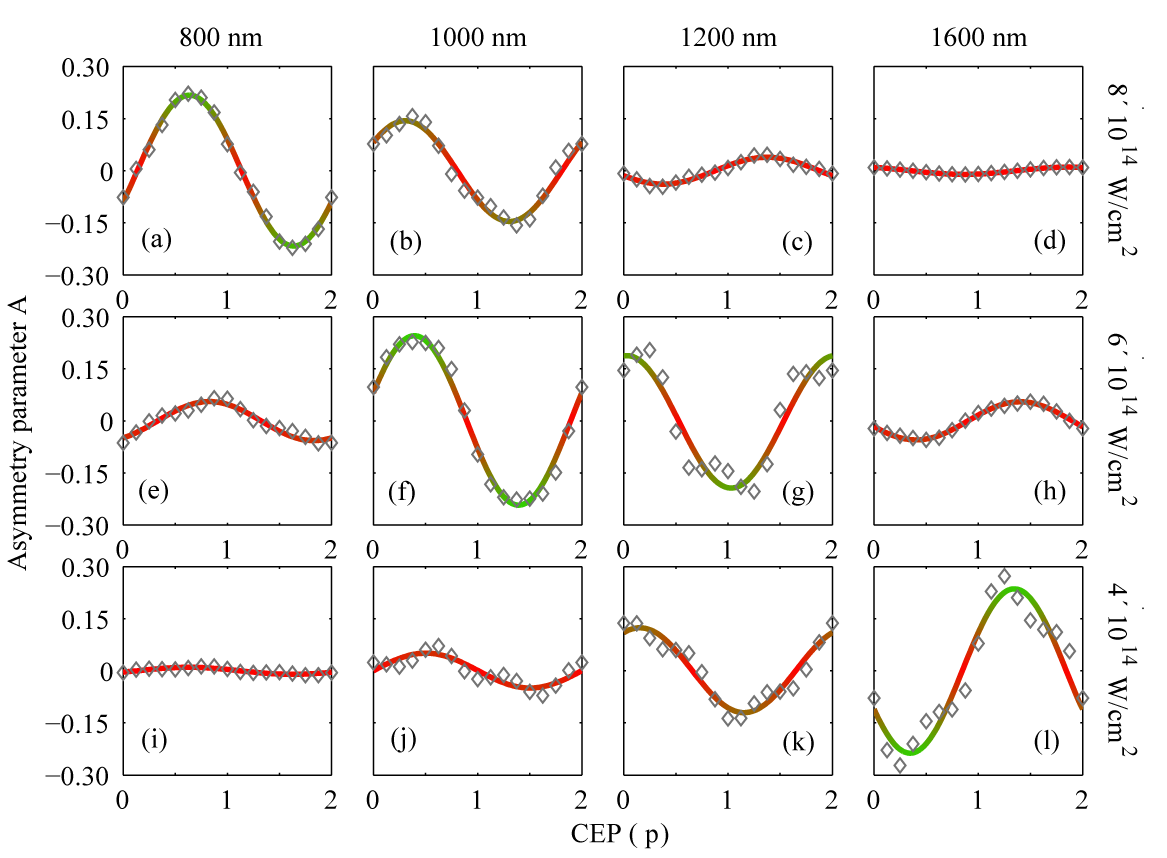}
\caption{\label{fig2} The dependence of the phase-sensitive asymmetry parameter $A$ on the wavelengths of 800, 1000, 1200 and 1600 nm (in increasing order from the left to the right column) for three different intensities of $8\times 10^{14}$ (top row), $6\times 10^{14}$ (middle row) and $4\times 10^{14}\ \mathrm{W/cm}^2$ (bottom row).}
\end{figure}

We aim to study the dependence of the asymmetric dissociation of H$_2^+$ on the wavelength with an intense two-cycle (FWHM) laser pulse.
Let us first define the absolute asymmetry of the dissociation as
$A=P_+-P_-$ \cite{Calvert}.
Here, the absolute asymmetry parameter $A$ is utilized because the total ionization and dissociation probabilities are comparable in our simulation and we attempt to find out the maximum probabilities of both the dissociation and electron localization.
Therefore, in the present work this parameter $A$ gives a more transparent representation of the asymmetric dissociation, whereas a normalized parameter could provide a misleading indication; for example, a large normalized asymmetry could be obtained in a scenario where the total dissociation probability is negligible.

Figure~\ref{fig2} reveals the dependence of the CEP-sensitive asymmetry parameter $A$ (diamonds) on the wavelength for three intensities of $8\times 10^{14}$, $6\times 10^{14}$ and $4\times 10^{14}\ \mathrm{W/cm}^2$, respectively.
On the one hand, the amplitude of the asymmetry is changed significantly with the wavelength. For higher intensity (top row), the maximum dissociation asymmetry $A_{max}$ decreases with the wavelength, while for lower intensity (bottom row), we observe an opposite variation of $A_{max}$.
Here, $A_{max}$ is defined as the maximum of $|A|$ in each panel. For the intermediate intensity (middle row), $A_{max}$ first increases and then decreases with the wavelength.
On the other hand, the data of the asymmetry parameter $A$ in Fig.~\ref{fig2} are fitted to sine functions (thick curves). We find that there is a phase-shift of the sine function between the panels. This means that $A_{max}$ appears at different values of the CEP when the wavelength or intensity is changed.

\begin{figure}
\centering\includegraphics[width=11cm]{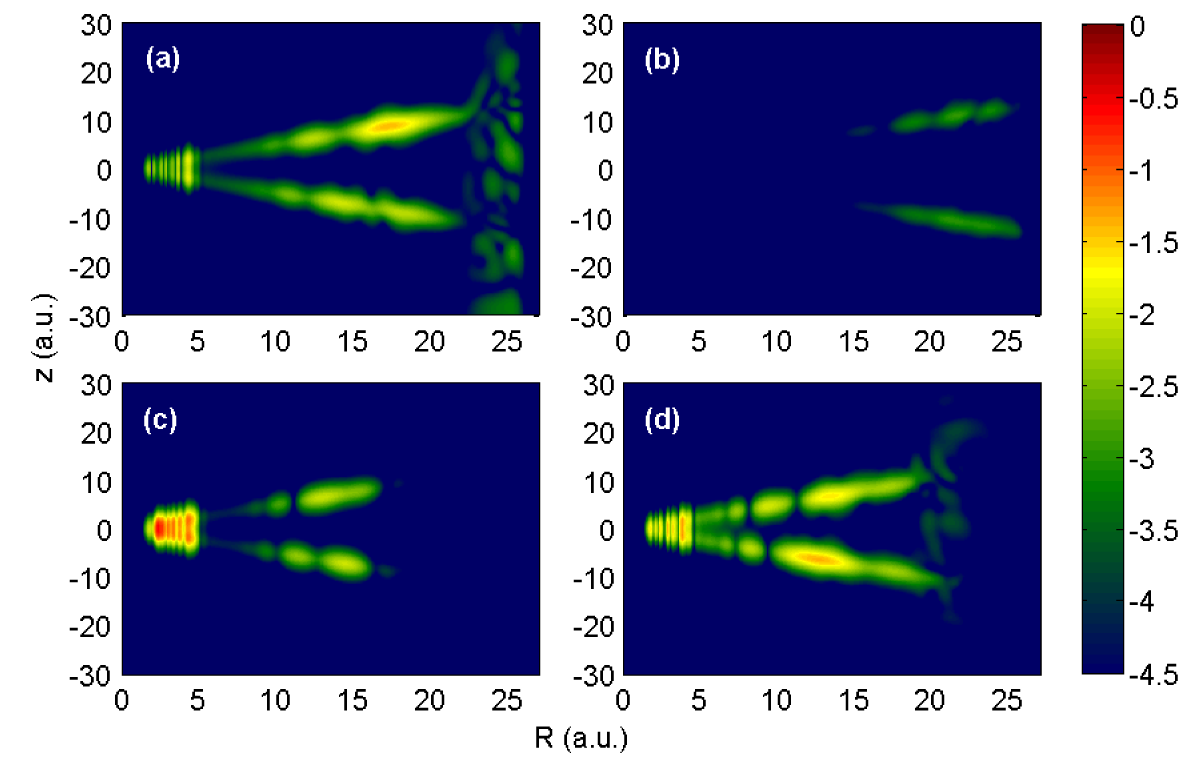}
\caption{\label{fig3} (Multimedia online) Snapshots of H$_2^+$ electron-nuclear probability density distribution taken at $t=30 \ \mathrm{fs}$ for four cases, in which the laser parameters for (a) (\textcolor[rgb]{0.00,0.00,1.00}{Media 1}), (b) (\textcolor[rgb]{0.00,0.00,1.00}{Media 2}), (c) (\textcolor[rgb]{0.00,0.00,1.00}{Media 3}) and (d) (\textcolor[rgb]{0.00,0.00,1.00}{Media 4}) are the same as those in Figs.~\ref{fig2}(a),~\ref{fig2}(d),~\ref{fig2}(i) and~\ref{fig2}(l), respectively, except that the CEP is fixed at $0.5\pi$. The color scale is logarithmic.}
\end{figure}

In order to gain an intuitive understanding of the electron localization in the dissociation of H$_2^+$, we simulate the real-time propagation of the electron-nuclear wavepackets in the external fields of four cases, as shown in Figs.~\ref{fig3}(a) to~\ref{fig3}(d) (Media 1--4), respectively.
For the four cases, the CEP of the pulse is fixed at $0.5\pi$, and the other laser parameters are the same as those in Figs.~\ref{fig2}(a),~\ref{fig2}(d),~\ref{fig2}(i) and~\ref{fig2}(l), respectively.
By following the evolution of the probability density in the external fields (Media 1--4), a strong vibration of the probability density can be observed when the pulse is applied. Meanwhile, part of wavepackets are ionized in the $\pm z$ direction at the extrema of the electric field. Apparently, the ionized probability density for higher intensity is much more than that for lower intensity.
At the tail of the pulse, the ionization almost stops. Part of the vibrational wavepackets dissociate slowly along the upper or lower proton, and the rest stay in the bound states.

According to the real-time evolution of wavepackets in the interaction, the dependence of $A_{max}$ on the wavelength for the relative higher and lower intensities in Fig.~\ref{fig2} can be qualitatively explained as follows. In the case of $8\times10^{14}\ \mathrm{W/cm}^2$, a great deal of the wavepackets are ionized.
For 800 nm, most of the wavepackets dissociate with the protons due to the nuclear vibration after the ionization [Fig.~\ref{fig3}(a), \textcolor[rgb]{0.00,0.00,1.00}{Media 1}].
But longer wavelength will cause stronger vibration and thus much more wavepackets will be further ionized by the driving field of high intensity \cite{Liucd}.
Consequently, for 1600 nm, almost all the wavepackets have been ionized before the pulse is turned off and only few wavepackets are dissociated [Fig.~\ref{fig3}(b), $\mathrm{\textcolor[rgb]{0.00,0.00,1.00}{Media\ 2}}$].
Note that the absolute asymmetry parameter $A$ is closely associated with the total dissociation probability \cite{Calvert}.
Thus, under the intensity of $8\times 10^{14}\ \mathrm{W/cm}^2$, the reduce of the dissociation probability leads to the decrease of $A_{max}$ with the wavelength.
However, the situation is quite different in the case of relative lower pulse intensity, where few wavepackets are ionized due to the relative low intensity. For 800 nm, because of the weak nuclear vibration, there are few dissociative wavepackets and most of the wavepackets stay in the bound states [Fig.~\ref{fig3}(c), \textcolor[rgb]{0.00,0.00,1.00}{Media 3}].
But the nuclei vibrate more and more intensely with increasing wavelength. For 1600 nm, only a few wavepackets remain in the bound states. Most of them are dissociated and localized on the dissociating protons [Fig.~\ref{fig3}(d), \textcolor[rgb]{0.00,0.00,1.00}{Media 4}].
As a result, $A_{max}$ increases with the wavelength for the intensity of $4\times10^{14}\ \mathrm{W/cm}^2$.

\begin{figure}
\centering\includegraphics[width=11cm]{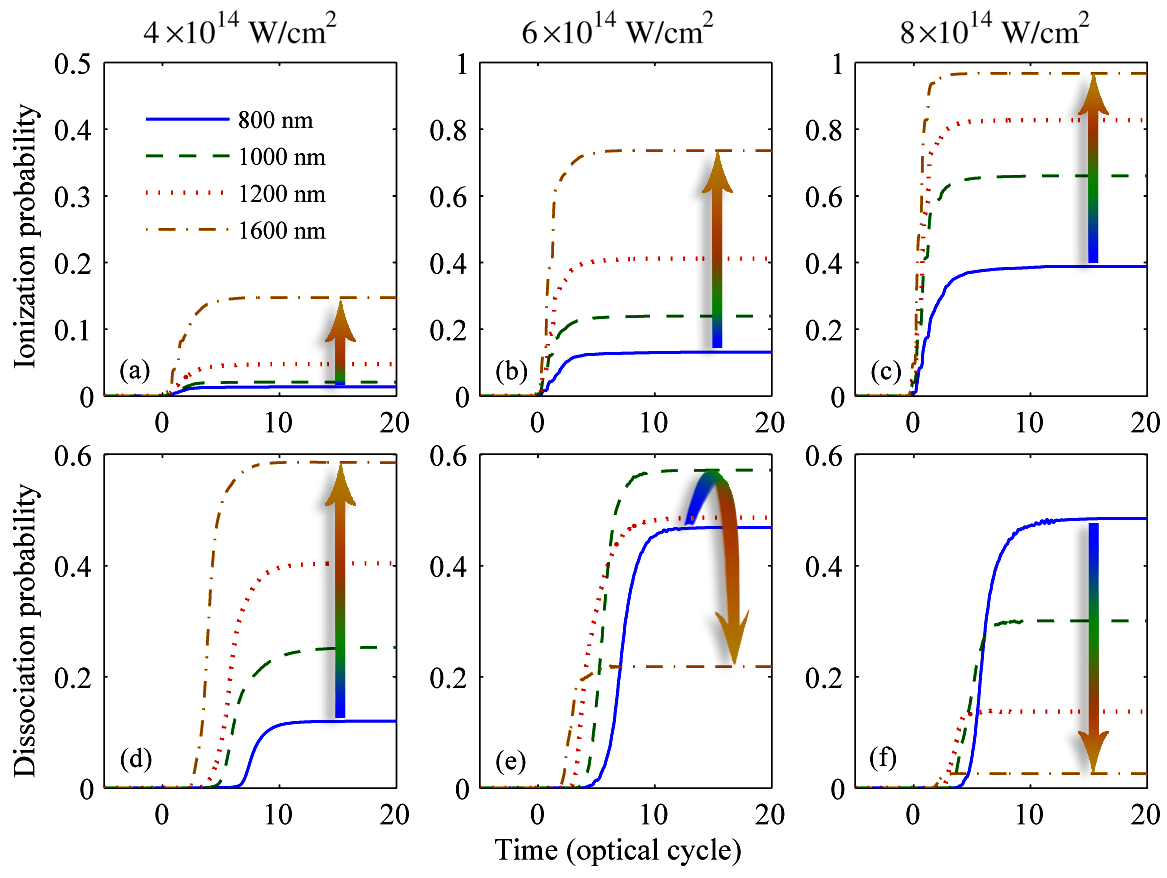}
\caption{\label{fig4} Time-dependent ionization and dissociation probabilities under four wavelengths for peak intensities of $4\times 10^{14}$ (left column), $6\times 10^{14}$ (middle column) and $8\times 10^{14}\ \mathrm{W/cm}^2$ (right column). The color arrows indicate the dependence of the final probabilities on the pulse wavelength.}
\end{figure}

To gain further insight in the relation between the ionization and dissociation that influences the maximum asymmetry of electron localization,
the time-dependent ionization and dissociation probabilities during the interactions are calculated. The results under four wavelengths for different intensities are depicted in Fig.~\ref{fig4}.
Firstly, in the left column of Fig.~\ref{fig4}, the total ionization probabilities for all wavelengths are less than 15\%, and the total dissociation probability increases rapidly with the wavelength.
However, when the intensity increases, the ionization increases more rapidly for longer wavelength, as shown in the middle column of Fig.~\ref{fig4}.
More than 40\% of the wavepackets are ionized for $1200\ \mathrm{nm}$ and $1600\ \mathrm{nm}$,
which directly reduce the total dissociation probabilities.
Lastly, the ionization are further enhanced for the intensity of $8\times 10^{14}\ \mathrm{W/cm}^2$, with the result that the total dissociation probability decreases with wavelength, as shown in the right column of Fig.~\ref{fig4}.
Note again that the asymmetry $A$ is closely related to the dissociation probability. Thus the variation of the total dissociation probability under different intensities discussed above leads to the corresponding dependence of $A_{max}$ on the wavelength in each row of Fig.~\ref{fig2}.

Now we discuss the wavelength effect on the electron localization in more detail
by comparing the results in Fig.~\ref{fig2} and Fig.~\ref{fig4}. On the one hand, as shown in Fig.~\ref{fig4}(e), the final dissociation probabilities for 800 nm (solid curve) and 1200 nm (dotted curve) are approximately equal.
However, the corresponding $A_{max}$ for 1200 nm in Fig.~\ref{fig2}(g) is larger than that for 800 nm in Fig.~\ref{fig2}(e).
On the other hand, although the dissociation probability for the wavelength of 800 nm under the intensity of $4\times 10^{14}\ \mathrm{W/cm}^2$ [solid curve in Fig.~\ref{fig4}(d)] is larger than that for 1600 nm under $8\times 10^{14}\ \mathrm{W/cm}^2$ [dash-dot curve in Fig.~\ref{fig4}(f)], the $A_{max}$ in the two cases are almost the same, as shown in Figs.~\ref{fig2}(i) and~\ref{fig2}(d).
In addition, some similar situations can be also found for other laser parameters. These features show that the longer-wavelength pulses are more effective to create stronger electron localization asymmetry.

\begin{figure}
\centering\includegraphics[width=8.5cm]{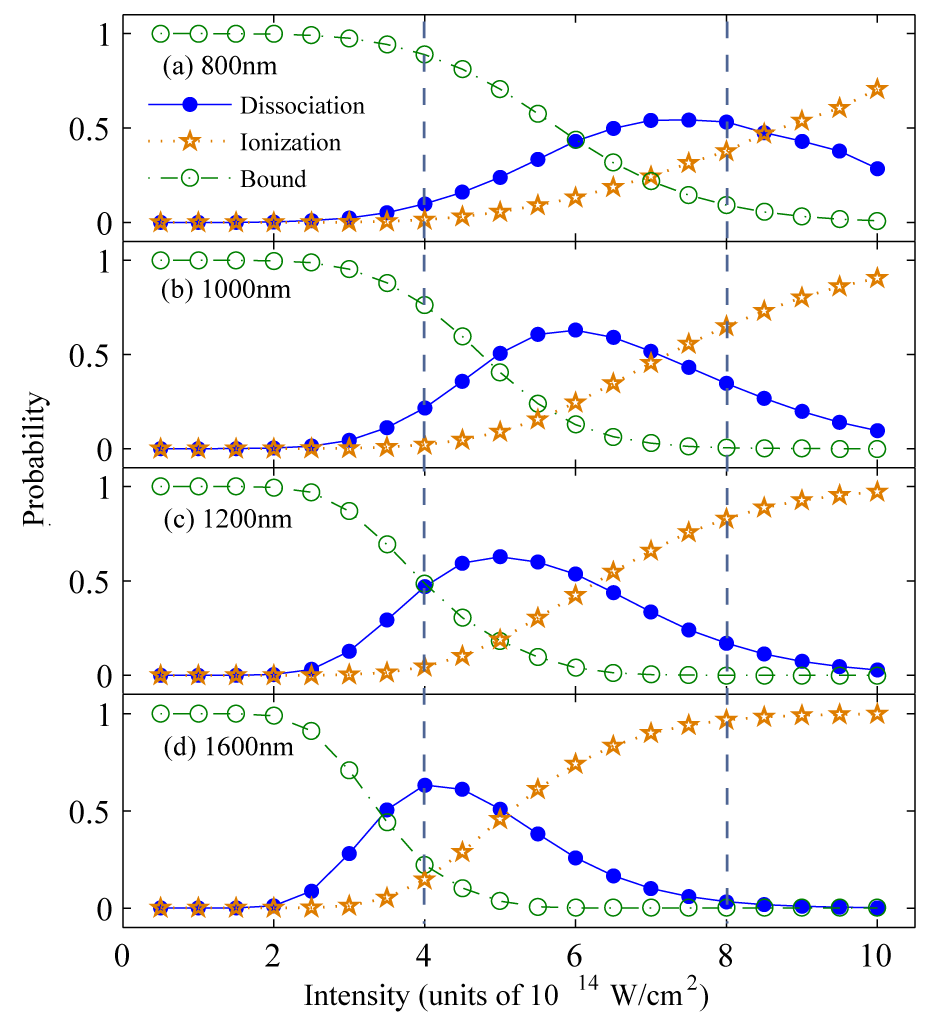}
\caption{\label{fig5} The final dissociation (dots), ionization (stars) and survival (rings) probabilities as a function of the intensity for the wavelengths of (a) 800, (b) 1000, (c) 1200 and (d)  $1600\ \mathrm{nm}$. The CEP is fixed at $0.5\pi$.}
\end{figure}

For comprehensive understanding of the laser-molecular interaction under different pulse wavelengths, we show the dissociation, ionization and bound probabilities as a function of the pulse intensity varying from $5\times 10^{13}$ to $1\times 10^{15}\ \mathrm{W/cm}^2$ for the wavelengths of 800, 1000, 1200 and 1600 nm in Figs.~\ref{fig5}(a) to~\ref{fig5}(d), respectively. The CEP of the pulse is fixed at $0.5\pi$ and the results do not vary significantly with the CEP in our calculation.
Obviously, there is a peak value of the dissociation probability for each wavelength in Fig.~\ref{fig5}, and its position moves from high to low intensities when the wavelength is increased. The offset of the dissociation peak can be understood as that the longer-wavelength pulse can create stronger nuclear vibration and thus the dissociation probability increases more rapidly.
But meanwhile, because of the intense nuclear vibration for the long-wavelength pulse, the ionization also increases rapidly, resulting in the early decrease of the dissociation probability.

By focusing on the given region of the intensity between the vertical dash lines in Fig.~\ref{fig5}, one can find that the variations of the dissociation probability with the intensity under different wavelength are generally coincident with those of $A_{max}$ in each column of Fig.~\ref{fig2}, respectively.
For the laser parameters of Figs.~\ref{fig2}(a),~\ref{fig2}(f) and~\ref{fig2}(l), strong asymmetries are obtained, approximately corresponding to the dissociation peaks in Fig.~\ref{fig5}.
However, when the dissociation reaches its peak value, the ionization probability is quite high for 800 nm and becomes lower for longer wavelength.
For the case of 1600 nm and
$4\times 10^{14}\ \mathrm{W/cm}^2$ in Fig.~\ref{fig5}(d), the total dissociation and ionization probabilities are 0.60 and 0.14, respectively.
By varying the CEP of the pulse, the $A_{max}$ shown in Fig.~\ref{fig2}(l) are as high as 0.273,
where the electron localization probabilities $P_+$ and $P_-$ are 0.436 and 0.163, respectively. In this scenario, more than 72\% of the dissociative (and 43\% of all) wavepackets are localized on one of the dissociating nuclei.
This indicates that the mid-infrared pulse can effectively localize the electron at a chosen nucleus with weak ionization.

\begin{figure}
\centering\includegraphics[width=12.8cm]{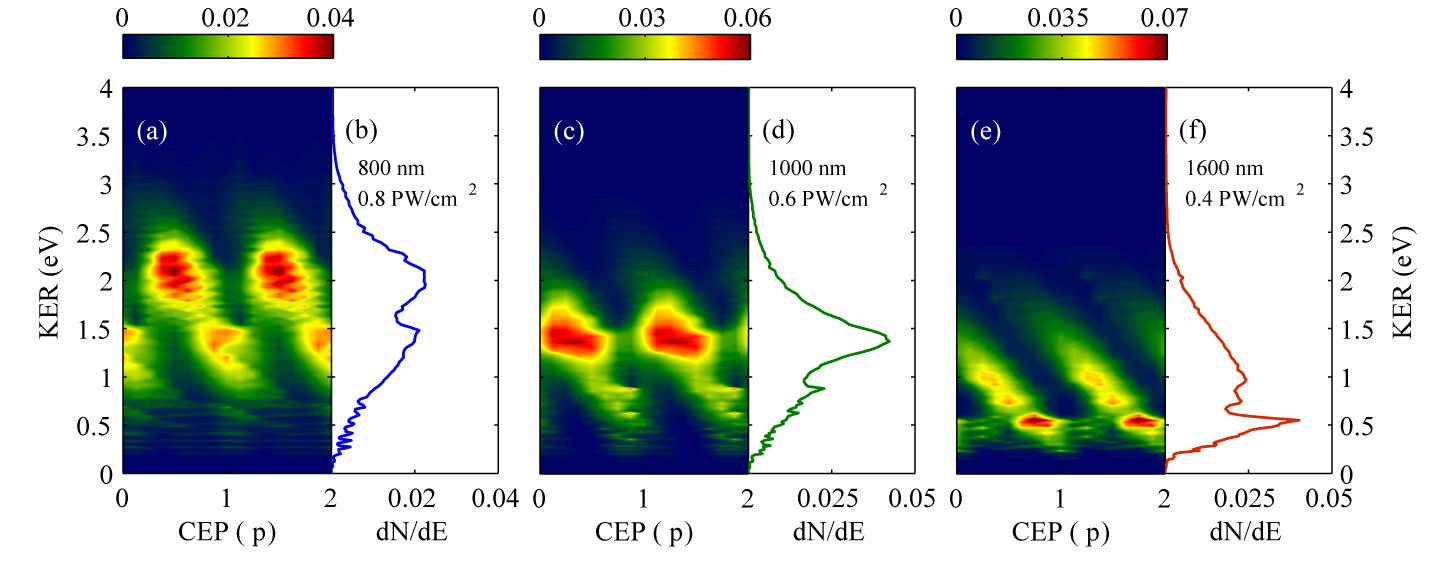}
\caption{\label{fig6} CEP-dependent and CEP-averaged KER spectra for different combinations of the pulse wavelength and intensity. The laser parameters for (a), (c) and (e) are the same as those in Figs.~\ref{fig2}(a), (f) and (l), respectively.}
\end{figure}

Next, we turn to the physical mechanism responsible for the phase-shift of the asymmetry $A$ in Fig.~\ref{fig2}. First of all, to obtain more detailed insight into the CEP control of electron localization, we calculate the fragment kinetic energy release (KER) distribution using the `virtual detector' method \cite{Feuerstein}.
Figures~\ref{fig6}(a),~\ref{fig6}(c) and~\ref{fig6}(e) show the CEP-dependent KER distributions for different combinations of the pulse wavelength and intensity, respectively. The solid curves in the Figs.~\ref{fig6}(b),~\ref{fig6}(d) and~\ref{fig6}(f) indicate the corresponding CEP-averaged KER spectra. As shown in Fig.~\ref{fig6}, the KER distributions peak at around 2 eV, 1.4 eV and 0.6 eV for the 800-nm, 1000-nm and 1600-nm pulses, respectively, indicating that the above-threshold dissociation contributes significantly to the KER spectra. The contributions from bond-softening are relatively weak
due to the high pulse intensity ($\geq4 \times 10^{14}\ \mathrm{W/cm}^2$) \cite{Roudnev2}.

According to the KER spectra in Fig.~\ref{fig6}, we propose a possible simplified explanation for the phase-shift of the asymmetry.
As illustrated in Fig.~\ref{fig1}(b), the wavepackets are bound in the initial state at first and begin to vibrate after the pulse is applied. Then some vibrational wavepackets on the $1s\sigma_g$ state absorb three photons and transfer to the dissociative $2p\sigma_u$ state. When the wavepackets on both states pass through the one-photon crossing point, the electric field couples the two states and then population is transferred between them. The coherent superposition of the dissociative wavepackets on the gerade and ungerade states finally leads to a localization of the electron at either of the nuclei \cite{Kremer,He2,Ray}.
Because (a) the motion of the wavepackets depends on the pulse intensity and wavelength, (b) the one-photon crossing point depends on the wavelength, and (c) the coupling field is sensitive to the CEP, the population transfer will be different if the intensity, wavelength or CEP is changed. As a result, the electron localization depends on
the CEP, intensity and wavelength of the pulse.

\begin{figure}
\centering\includegraphics[width=12cm]{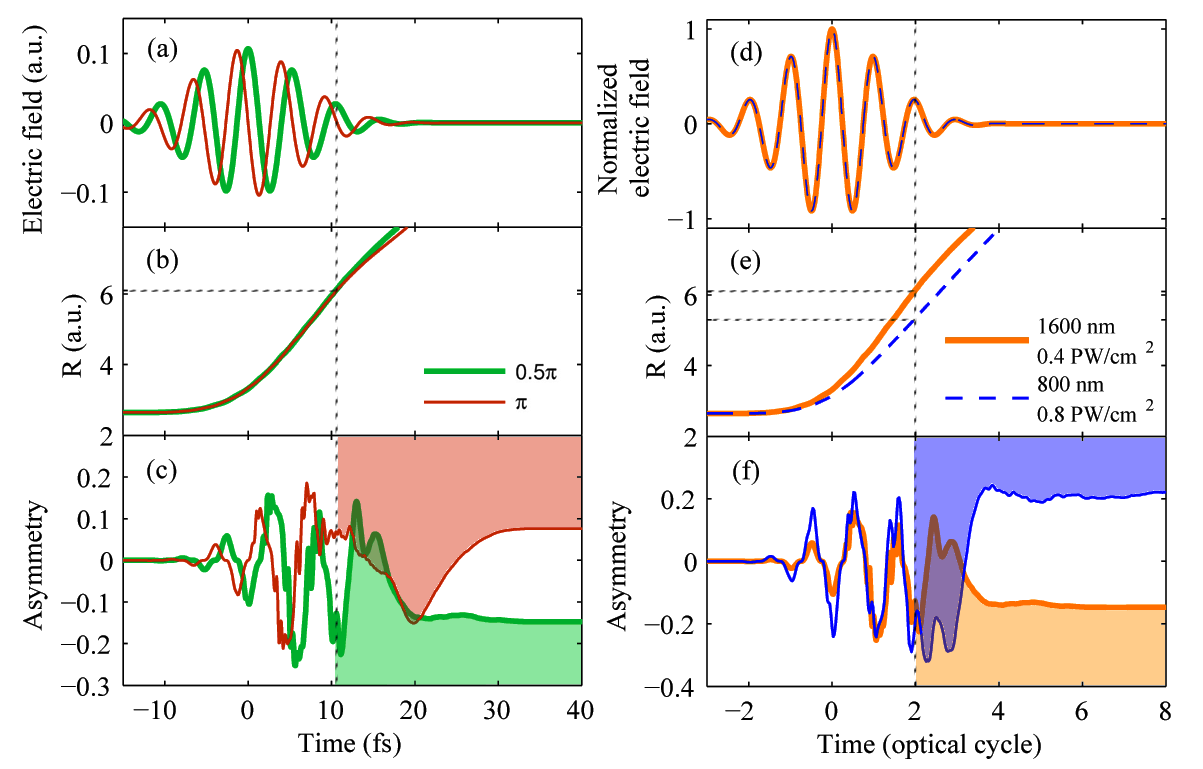}
\caption{\label{fig7} The electric fields, internuclear distances and asymmetries as a function of time (in different units for the left and right columns). For the left column, the pulse parameters are with the same wavelength of 1600 nm and intensity of $4\times10^{14}\ \mathrm{W/cm}^2$ but different values of the CEP. For the right column, the pulse parameters are with the same CEP of $0.5\pi$ but different wavelengths and intensities. The horizontal and vertical dash lines indicate the one-photon coupling points (about 6.1 a.u. for 1600 nm and 5.2 a.u. for 800 nm) and the time when the average internuclear distances reach the points, respectively.}
\end{figure}

To verify the interpretation for the origin of the phase-shift discussing above, we show a comparison of the time-dependent asymmetry versus the time-dependent internuclear distance for different laser parameters in Fig.~\ref{fig7}. Here are the calculation details.
For the TDSE approach beyond Born-Oppenheimer approximation in the present study, the individual time-dependent population on the ground or excited states couldn't be calculated in principle. Instead, we can extract the time-dependent bound wavepackets by
\begin{eqnarray}
\Psi_{\mathrm{bound}}(R,z;t)=\sum_{n=0}^{N}\langle v_n(R,z)|\Psi(R,z;t)\rangle v_n(R,z),
\end{eqnarray}
with $v_n$ the vibrational bound states and $N=21$ the number of $v_n$ in our model.
Then the time-dependent probabilities of the dissociative wavepackets localized on the upper or lower nucleus can be estimated by
\begin{eqnarray}
P'_{\pm}(t)&=&\int_0^{10}\mathrm{d}R\int_{0}^{\pm 10} \mathrm{d}z\ |\Psi(R,z;t)-\Psi_{\mathrm{bound}}(R,z;t)|^2 \nonumber \\
 & & +\int_{10}^{R_{max}}\mathrm{d}R\int_{\pm \frac{R}{2}-5}^{\pm \frac{R}{2}+5} \mathrm{d}z\  |\Psi(R,z;t)-\Psi_{\mathrm{bound}}(R,z;t)|^2,
\end{eqnarray}
where $R_{max}$ corresponds to the boundary of the grid.
The time-dependent expectation values of the internuclear distance $\overline{R(t)}$ for the total wavepackets is given by \cite{Steeg}
\begin{eqnarray}
\overline{R(t)}=\frac{\int _0^{R_{max}}\mathrm{d}R \int _{-z_{max}}^{z_{max}}\mathrm{d}z\ R|\Psi(R,z;t)|^2}{\int _0^{R_{max}}\mathrm{d}R \int _{-z_{max}}^{z_{max}}\mathrm{d}z\ |\Psi(R,z;t)|^2}
\end{eqnarray}
with $\pm z_{max}$ the boundaries of the grid.

We firstly focus on the situation for different values of CEP and otherwise identical laser parameters, as shown in the left column of Fig.~\ref{fig7}. The variations of the time-dependent internuclear distances are almost coincident before they reach the coupling point, and the asymmetries [$P'_+(t)-P'_-(t)$] oscillate with the electric fields in a similar way. However, after the internuclear distances reach the coupling point, the asymmetries vary in completely different ways and finally lead to the different localizations of the electron, as shown in Fig.~\ref{fig7}(c).
Then we turn to the situation for the same CEP but different wavelengths and intensities, as shown the right column of Fig.~\ref{fig7}.
The optical cycle (o.c.) has been used as the unit of time and the intensities of the pulses have been normalized so that the electric fields share the same profile. In Fig.~\ref{fig7}(e), the internuclear distances reach their coupling points approximately at $t=2$ o.c. (but actually not the same time).
Before $t=2$ o.c., both the asymmetries in Fig.~\ref{fig7}(f) oscillate with the pulse almost in a same way. However, because of the different molecular motions as well as photon energies, the asymmetries change quite differently after the coupling point, resulting in the electron localization on different nuclei.
Overall, the two situations discussing above indicate that the one-photon coupling point plays an important role in determining the asymmetry, and the asymmetry depends not only on the CEP but also the intensity and wavelength of the pulse even though the electric field share the same profile.

\section{Conclusion}
In conclusion, we have studied the wavelength dependence of the asymmetric dissociation of H$_2^+$ by solving numerically the non-Born-Oppenheimer time-dependent Schr$\ddot{\mathrm{o}}$dinger equation.
By varying the wavelength, significant changes of the electron localization asymmetry have been observed. The strong asymmetries that are obtained at different intensities are associated with the offset of the dissociation peak, and the phase-shift of the asymmetry is attributed to the different population transfers by the remaining fields after the internuclear distances reach the one-photon coupling point.
Moreover, a high degree of control over the electron localization on a chosen nucleus with relative low ionization probability of the molecule has been achieved with the 1600-nm pulse, suggesting that the mid-infrared few-cycle pulse can be used to effectively control the electron localization in molecular dissociation with weak ionization.

\section*{Acknowledgement}
This work was supported by the National Natural Science Foundation
of China under Grants No. 60925021, 10904045, 10734080 and the National
Basic Research Program of China under Grant No. 2011CB808103. This
work was partially supported by the State Key Laboratory of
Precision Spectroscopy of East China Normal University.

\end{document}